\documentclass[conference,letterpaper]{IEEEtran}

\IEEEoverridecommandlockouts
\usepackage{amsmath} 
\usepackage{amssymb} 
\usepackage{MnSymbol}
\usepackage{extarrows}
\usepackage{enumerate} 
\usepackage{bm} 
\usepackage{booktabs}
\usepackage{array}

\usepackage{graphicx} 
\usepackage{subfigure} 
\usepackage{multirow} 

\usepackage{cite}
\usepackage{amsmath,amsfonts}
\usepackage{algorithmic}
\usepackage{array}
\usepackage[caption=false,font=normalsize,labelfont=sf,textfont=sf]{subfig}
\usepackage{textcomp}
\usepackage{stfloats}
\usepackage{url}
\usepackage{verbatim}
\usepackage{xcolor}
\usepackage{graphicx}
\newtheorem{theorem}{Theorem}[section]
\newtheorem{definition}{Definition}[section]

\newtheorem{lemma}{Lemma}[section]

\newtheorem{proposition}{Proposition}[section]
\newtheorem{corollary}{Corollary}[section]
\newtheorem{remark}{Remark}[section]

\newcommand{\pf}{\textbf{Proof: }}
\newcommand{\e}{\hfill$\blacksquare$}

\newcommand{\R}{\mathbb{R}}

\usepackage{epstopdf}

\allowdisplaybreaks[4] 

\date{}
\def\BibTeX{{\rm B\kern-.05em{\sc i\kern-.025em b}\kern-.08em
		T\kern-.1667em\lower.7ex\hbox{E}\kern-.125emX}}
\usepackage{balance}
\begin{document}

	\title{New Partial Orders of Polar Codes for BMSC
	}
	
%
%
%

 \author{%
	\IEEEauthorblockN{Liuquan Yao\IEEEauthorrefmark{1},
	Zhichao Liu\IEEEauthorrefmark{1},
		Yuan Li\IEEEauthorrefmark{2},
		Huazi Zhang\IEEEauthorrefmark{2},
	Jun Wang\IEEEauthorrefmark{2}
Guiying Yan\IEEEauthorrefmark{3}
and Zhi-Ming Ma\IEEEauthorrefmark{3}}
	\IEEEauthorblockA{\IEEEauthorrefmark{1}%
		University of Chinese Academy and Sciences, Beijing, China,
		\{yaoliuquan20, liuzhichao20\}@mails.ucas.ac.cn}
	\IEEEauthorblockA{\IEEEauthorrefmark{2}%
		Huawei Technologies Co. Ltd., China,
		\{liyuan299, zhanghuazi, justin.wangjun\}@huawei.com}
	\IEEEauthorblockA{\IEEEauthorrefmark{3}%
		Academy of Mathematics and Systems Science, CAS, Beijing, China,
	yangy@amss.ac.cn, 	mazm@amt.ac.cn}
}

	\maketitle
	
	\begin{abstract}
		In this paper, we define partial orders (POs) of polar codes based on the Bhattacharyya parameter and the bit-error probability, respectively. These POs are applicable to arbitrary binary memoryless symmetric channel (BMSC). Leveraging the extremal inequalities of polarization transformation, we derive new POs for BMSC based on the corresponding POs observed in the Binary Erasure Channel (BEC). 
		We provide examples that demonstrate the inability of existing POs to deduce these novel POs. Furthermore, we establish upper bounds for the expansion parameter $\beta$ if the polar codes constructed by   $\beta$-expansion method obey these POs.
	\end{abstract}
	\section{Introduction}\label{section1}
	Polar codes, invented by Ar{\i}kan \cite{arikan2008}, have been proven to achieve the capacity in any BMSC with low-complexity encoding and decoding algorithms. In polar codes,  $N$ independent channels $W$ are transformed to $N$ synthesized channels $W_N^{(i)}, i=0,1,\cdots, N-1$. As the code length $N$ tends to infinity,   the fraction of almost noiseless channels approaches the symmetric capacity of $W$, while the remaining channels become almost pure-noise. Thanks to the polarization phenomenon, we choose the best $K$ synthesized channels to transmit information while the remaining channels are frozen to zeros. The information bits can be decoding successively using the successive cancellation (SC)   algorithm with a complexity of $O(N\log N)$. 
	
	To identify reliable synthesized channels, efficient methods such as   density evolution   algorithm \cite{densityTanaka}, Gaussian approximation   algorithm \cite{b11}, and the channel-independent PW construction method \cite{beta-expansion} have been proposed. The quality of synthesized channels $W_N^{(i)}$ can be evaluated by various parameters, such as the capacity, the Bhattacharyya parameter or the bit-error probability. However, as the code length $N$ increases, calculating these parameters for all the synthesized channels becomes rather challenging. Consequently, fast algorithms for polar code construction have been developed. For instance, merging algorithms based on channel degradation, which provide tightened bounds for the Bhattacharyya parameter and the bit-error probability have been established in  \cite{2011How} and \cite{2021greedy}.	
	
POs between synthesized channels $W_N^{(i)}$  have been investigated in several papers  \cite{01<10, 2016PO, Wu2019, lin2023optimal, wang2023density, wang2023fast}. The primary objective of PO research is to identify a universal (i.e., channel-independent) comparison method for synthesized channels, aiming to reduce the complexity of polar code construction. In \cite{01<10}, the authors defined POs on BMSCs based on channel degradation and \cite{2016PO} proposed a simplified algorithm for polar code construction based on POs, which only requires 2.8\% channel evaluations comparing with the algorithm in  \cite{2011How} for  a specific set of parameters ($N=2^{20}$ and AWGN channel with noise variance $\sigma^2=0.631$). The utilization of POs eliminates unnecessary calculations. For example, in \cite{2016PO},  approximately 23.06\% of the synthesized channels were found to be superior to $W_N^{(436586)}$, while 23.07\% of the synthesized channels are worse, which is independent of the initial channel $W$. This observation implies that if $W_N^{(436586)}$ is selected as an information bit, around 23.06\% of the synthesized channels can also be chosen as information bits. Conversely, if $W_N^{(436586)}$ is frozen, approximately 23.07\% of the positions can be frozen as well.
	
	Although POs show promise in reducing the complexity of polar code construction, there has been limited research about POs on BMSCs. One fundamental PO is channel degradation, which has two well-established conclusions: 1) $0\preceq 1$, i.e., performing up polarization transform is always worse than down polarization transform, 2) $01\preceq 10$, \textit{i.e.} performing up polarization followed by down polarization is always worse than the converse order. For the definitions of up and down polarization, refer to  \cite{arikan2008}. However, beyond these conclusions, there is a lack of research on POs for BMSC. In contrast, there are more POs  in the BEC. In \cite{Wu2019}, the authors proposed channel degradation between two synthesized channels for BMSC under specific conditions and provided a simplified form when the initial channel is a BEC. In \cite{lin2023optimal},   Green theorem was employed to confirm a conjecture presented in \cite{Wu2019}. Additionally, \cite{wang2023density} introduced an algorithm to verify the existence of a PO between any two polarization paths under BEC. A compilation of existing POs for BEC was provided in \cite{wang2023fast}.
	
	In this paper, we introduce novel POs for BMSCs. Our approach leverages the existing POs for BEC as building blocks to derive new POs applicable to arbitrary BMSC. Our results encompass certain POs that cannot be deduced from known conclusions. Additionally, we establish upper bounds for the expansion parameter $\beta$ in order to make the  polar codes constructed by the $\beta$-expansion method comply with the newly discovered POs. A summary of our findings is presented in Table \ref{table 1}.
	
	%
	%
	
	\begin{table}[htbp]
		\caption{New POs for BMSC}
		\begin{center}
			\begin{tabular}{|c|c|c|c|}
				\hline
				&	\textbf{Conditions}&\textbf{POs}&\textbf{Theorem}\\
				\cline{1-4} 
				1&	$\gamma  \preceq_{BEC}  \alpha$&$\gamma1\preceq_Z 1\alpha$& \ref{PO Z from cancel 1}\\
				\hline
				2&	$\gamma  \preceq_{BEC}  \alpha$&$\gamma0\preceq_P 1\alpha$& \ref{P1}\\
				\hline
				3& $\alpha\preceq_{BEC} 0^q0\tau$&$\alpha\preceq_P 0^q1\tau$& \ref{theorem of Pe}\\
				\hline
				4& $\tau_1\preceq \tau_2, \alpha \preceq_Z  \gamma$   &$\tau_1 \alpha 1^p\preceq_Z  \tau_2\gamma 1^p$& \ref{rule}\\
				\hline
				5& $\tau_1\preceq \tau_2, \alpha \preceq_P  \gamma$&$\tau_1 \alpha 0^p\preceq_P  \tau_2\gamma 0^p$& \ref{rule}\\
				\hline
				6&$0^p\alpha 1^r\preceq_Z 1^q\gamma 0^s$&$0^p\tau \alpha 1^t 1^r\preceq_Z 1^q \tau \gamma 1^t0^s$& \ref{rule}\\
				\hline
				7& $	0^p\alpha 1^r\preceq_P 1^q\gamma 0^s$&$0^p\tau \alpha 0^t 1^r\preceq_P 1^q \tau \gamma 0^t0^s$& \ref{rule}\\
				\hline
			\end{tabular}
			\label{table 1}
		\end{center}
	\end{table}
	
	This paper is organized as follows.  The definition of channel degradation and   known POs for BMSC are shown in Section \ref{section2}. Additionally, we provide an overview of related studies on POs   for BEC. In Section \ref{section3}, we introduce two types of new POs for BMSC based on the Bhattacharyya parameter and the bit-error probability, and show how to deduce the POs for BMSC from the known POs for BEC. At the end of Section \ref{section3}, we present examples that cannot be obtained using the known POs and utilize them to establish bounds for the expansion parameter $\beta$. Simulation results are shown in Section \ref{section4}. Finally, we draw some conclusions in Section \ref{section5}.

	\subsection*{Notations and definitions}
	
	Greek letters are used to represent polarization paths, with the exception of $\beta$   reserved for $\beta$-expansion. Paired vertical bars $|\cdot|$ denote the length of a path. We use the subscript $i$ to specify the  $i$-th polarization transform. $\bar{\alpha}$ denotes the bit inversion of $\alpha$.
	For example, $\alpha=001$ is the path that performs two up polarization transforms followed by one down polarization transform. Consequently, $|\alpha|=3$,  $\alpha_1=0, \alpha_2=0, \alpha_3=1$, $\overline{001}=110$. For convenience, we use $0^p1^q$ to represent the path consisting of $p$ times up polarization transforms followed by $q$ times down polarization transforms. We use $W^\alpha$  to denote the synthesized channel generated by the polarization path $\alpha$ with the initial channel $W$.
	
	We define  functions $Z_0(x):=1-(1-x)^2, Z_1(x):=x^2$ with $x\in [0,1]$, and then for path $\alpha$ with length $n$,  define
	\begin{equation}
	\begin{aligned}
	Z_\alpha(x)&=Z_{\alpha_1\alpha_2\cdots\alpha_n}(x)\\
	&:=Z_{\alpha_n}\circ Z_{\alpha_{n-1}}\circ\cdots\circ Z_{\alpha_1}(x),\;\; x\in[0,1].
	\end{aligned}
	\end{equation}
	For example, for any $p, q\in\mathbb{N}$,
	\begin{equation}
	Z_{{0^q1^p}}(x)=[1-(1-x)^{2^q}]^{2^p},\; Z_{1^p0^q}(x)=1-(1-x^{2^p})^{2^q}.
	\end{equation}
	\begin{remark}
		$Z_\alpha(x)$ is the Bhattacharyya parameter of $W^\alpha$ when $W$ is the BEC with the Bhattacharyya parameter $x$. $Z_\alpha$ is closely related to the capacity functions $I_\gamma$ defined in \cite{lin2023optimal}, in fact,
		\begin{equation}
		Z_\alpha(x)=I_{\bar{\alpha}}(x),\;\;\forall x\in[0,1].
		\end{equation}
	\end{remark}
	
	We use $I, Z, P_e$ to represent the symmetric capacity, the Bhattacharyya parameter and the bit-error probability respectively.
	
	\section{Partial Orders for BMSC and BEC}\label{section2}
	In this section, we review the known POs for BMSC. Additionally, recent results on POs for BEC are introduced.
	\subsection{Channel Degradation}
	Channel degradation is a fundamental criterion to compare two channels.
	\begin{definition}
		The channel $W$ is stochastically degraded  with respect to $Q$, if there exists a channel $P$,
		\begin{equation}
		W(y|x)=\sum_{z} Q(z|x)P(y|z),\;\;\forall x,y.
		\end{equation}
		Denote the channel degradation relationship by $W\preceq Q$.
	\end{definition}
	
	The notion of channel degradation has been studied in prior works such as \cite{Wu2019} and \cite{2008Modern}, serving as a fundamental PO. The significance of channel degradation is highlighted in the following proposition.
	\begin{proposition}\label{quanlities under degradation}
		\cite[Lemma 3]{Wu2019} If $W\preceq Q$, then
		\begin{equation}
		I(W)\le I(Q),\;\; P_e(W)\ge P_e(Q),\;\; Z(W)\ge Z(Q).
		\end{equation}
	\end{proposition}
	Proposition  \ref{quanlities under degradation} implies that the PO $W \preceq Q$ generated by channel degradation implies $Q$ is superior to $W$ in terms of channel parameters $I, Z$ and $P_e$. We summarize the known results regarding $\preceq$, primarily from \cite{01<10} and \cite{Wu2019}.
	\begin{proposition}\label{known PO in BMS}
		For arbitrary BMSCs $W$ and $Q$, we have:
		\begin{enumerate}
			\item $
			W^0\preceq W^1,\; W^{01}\preceq W^{10}.
			$
			
			\item 	If $W\preceq Q$, then
			$
			W^1\preceq Q^1,\;\; W^0\preceq Q^0.
			$
			
			\item If $W^{\alpha}\preceq  W^{\gamma}$, where $\alpha, \gamma$ have the same length, and exist $n\ge 1$ and a path string $\tau$ with length $k\ge 0$, such that $W^{\alpha \tau 1^n}\preceq  W^{\gamma \tau 0^n}$, then for any string $\eta$,
			\begin{equation*}
			W^{\alpha \tau\eta 1^n}\preceq  W^{\gamma \tau\eta 0^n}.
			\end{equation*}	
		\end{enumerate}

	\end{proposition}
	
	We conclude the above results that in the polarization paths, $1$ is better than $0$ and 1 on the left is better than 1 on the right.
	In \cite{wang2023density}, an algorithm to find all POs for BEC with path length $n\le 5$ were provided, and we   verify whether these POs hold in BMSC by Proposition \ref{known PO in BMS}.

	Specifically, when $n=3$, the POs for BEC are still true for BMSC, except the pair $(100,011)$.

	When $n \geq 4$,  $(1100,1011)$, $(1010,0111)$, $(1000,0110)$  and $(10010,01111)$ are incomparable through Proposition \ref{known PO in BMS}.  Due to the stringent nature and computational complexity of POs based on channel degradation,
	we seek POs that are easier to verify. In the subsequent sections, we will define POs using the  Bhattacharyya parameter and the bit-error probability and derive additional POs for BMSC based on these criteria. Importantly, the aforementioned path pairs can be compared and ordered under the proposed POs.

	To establish a total order for all polarization paths, a well-known PO called $\beta$-expansion \cite{beta-expansion} was proposed.
	\begin{definition}
		Given $\beta\in \R$, define
		$
		B(\alpha):=\sum_{i=1}^{n} \beta^{n-i}\alpha_i,
		$
		where $n=|\alpha|$ is the length of $\alpha$. Write $\alpha\preceq_\beta \gamma$ iff
		$
		B(\alpha)\le B(\gamma).
		$
	\end{definition}
	
	The definition of $\beta$-expansion may not possess a physical interpretation. The primary objective of $\beta$-expansion is to identify an optimal value for $\beta$ that brings the partial order $\preceq_\beta$ significantly closer to certain physically meaningful POs.

	\subsection{Partial Orders in  BEC}
	BEC is the simplest BMSC, we first  summarize some properties of BEC. 
	\begin{proposition}
		If $W,Q$ are two BECs, then the following statements are equivalent.
		\begin{enumerate}
			\item $W\preceq Q$.
			\item $I(W)\le I(Q)$.
			\item $Z(W)\ge Z(Q)$.
			\item $P_e(W)\ge P_e(Q)$.
		\end{enumerate}
	\end{proposition}
	Therefore, we can define the PO just based on the Bhattacharyya parameter.
	\begin{definition}
		We write $\alpha\preceq_{BEC} \gamma$ iff
		\begin{equation}
		Z_{\gamma}(x)\le Z_{\alpha }(x),\;\;\forall x\in [0,1].
		\end{equation}
	\end{definition}
	Clearly, $\preceq_{BEC}$ defines a PO for polarization paths, and $\alpha\preceq_{BEC} \gamma$ means that for any initial channel $W$ which is a BEC, we have
	\begin{equation}
	W^\alpha \preceq W^\gamma ,\;\; Z(W^\gamma)\le Z(W^\alpha),
	\end{equation}
	and
	\begin{equation}
	I(W^\alpha)\le I(W^\gamma),\;\; P_e(W^\alpha)\ge P_e(W^\gamma).
	\end{equation}
	Note that the function $Z_{\alpha}(x)$ is a polynomial related to $\alpha$, so the efficient mathematical methods, such as Taylor-formula and root-isolation algorithms, can be used to verify the PO $\preceq_{BEC}$. We summary some results from \cite{lin2023optimal} and \cite{wang2023fast}.
	\begin{proposition}\label{BEC res}
		\begin{enumerate}
			\item $010^k10\preceq_{BEC} 100^k01, \forall k\in\mathbb{N}$.
			\item $1^m0^n\preceq_{BEC} 0^m1^n$ iff $(1-1/2^{2^m})^{2^n}\le 1/2.$
		\end{enumerate}
	\end{proposition}
	Moreover, there are other POs defined for BEC, such as Bernstein Bases PO, Average PO, Halfway-point PO,and refer to \cite{wang2023fast} for details.

	\section{New Partial Orders for BMSC}\label{section3}

	Due to the difficulty in verifying POs based on channel degradation, we propose the following two related POs:
	
	\begin{definition}
		We write $W\preceq_Z Q$, if $Z(W)\ge Z(Q)$; 	We write $W\preceq_P Q$, if $P_e(W)\ge P_e(Q)$.
		
		If $W^\alpha\preceq_Z W^\gamma$ for any BMSC   $W$, 	We write $\alpha\preceq_Z \gamma$; similarly, if $W^\alpha\preceq_P W^\gamma$ for any BMSC   $W$, 	We write $\alpha\preceq_P \gamma$.
	\end{definition}
	Obviously, for any path strings $\alpha, \gamma$
	\begin{equation}
 \alpha \preceq_Z \gamma
\text{ implies } \alpha \preceq_{\text{BEC}} \gamma,
	\end{equation}
	\begin{equation}
	 \alpha \preceq_P \gamma
	\text{ implies } \alpha \preceq_{\text{BEC}} \gamma.
	\end{equation}

	The newly defined POs provide several advantages:
	\begin{itemize}
		\item  They relax the constraints imposed by the original PO $\preceq$ and are easier to verify.
		\item  Unlike the specific PO $\preceq_{BEC}$, they are applicable to any BMSC. This broader applicability allows for a more comprehensive analysis of polarization paths in different channel scenarios.
		\item  The newly defined POs have physical significance as they are related to important parameters of the synthesized channels. This physical interpretation adds meaningful insights into the comparison of polarization paths.
	\end{itemize}
	
	In addition to the functions $Z_0(x), Z_1(x)$, we define 
	\begin{equation}
	L_0(x)=x\sqrt{2-x^2}.
	\end{equation}
	
	These three functions are closely related with the Bhattacharyya parameter of BMSC due to the following proposition.
	\begin{proposition}\label{bound functions}
		\cite{Wu2019} Let $W$ be a BMSC  , then
		\begin{equation}\label{1}
		L_0(Z(W))\le Z(W^0)\le Z_0(Z(W)),
		\end{equation}
		\begin{equation} 
		Z(W^1)=Z_1(Z(W)).
		\end{equation}
	\end{proposition}
	
	
	Similarly, for the bit-error probability, we have several bounds from \cite{relaxed2017} that
	\begin{equation}
	P_e(W^0)=2P_e(W)-2P_e(W)^2,
	\end{equation}
	\begin{equation}
	2P_e(W)^2\le P_e(W^1)\le \frac{1}{2}Z(W)^2,
	\end{equation}
	and
	\begin{equation}\label{Z and Pe}
	1-\sqrt{1-Z(W)^2}\le 2P_e(W)\le Z(W).
	\end{equation}

	\subsection{POs from BEC to BMSC}
	In this part, we prove the results presented in Table \ref{table 1}. The proof is mainly based on the extreme inequalities \eqref{1}-\eqref{Z and Pe}. We confirm a PO $\alpha\preceq_Z(\preceq_P) \gamma$ if the lower bound of $Z(W^\alpha)(P_e(W^\alpha))$ is large than the upper bound of $Z(W^\gamma)(P_e(W^\gamma))$.   Firstly, we consider the special case where the polarization path has the form $0^n1^m$ or $1^m0^n$.
	
	\begin{lemma}\label{functions}
		For any $k>0$,
		\begin{equation}
		\begin{aligned}
		&\underbrace{L_0\circ L_0\circ\cdots\circ L_0}(x)=\sqrt{1-(1-x^2)^{2^{k }}}=:L_0^{(k)}(x).\\
		&\;\;\;\;\;\;\;\;\;\;\;\;\;\;\;k
		\end{aligned}
		\end{equation}
	\end{lemma}
	\pf 
	Clearly,
	\begin{align}
	&L_0^{(k)}\circ L_0^{(l)}(x)=\sqrt{1-[1-(1-(1-x^2)^{2^l})]^{2^k}}\\
	&=\sqrt{1-(1-x^2)^{2^{k+l}}}=L_0^{(k+l)}(x),\;\;\forall k,l>0.
	\end{align}
	\e
	
	%
	%
	%
	Lemma \ref{functions} provides a simplified lower bound on the Bhattacharyya parameter after $k$ times up polarization transforms. Consequently, we obtain the upper and lower bounds for the Bhattacharyya parameter of the paths $1^m0^n$ and $0^m1^n$ with the initial channel $W$.
	\begin{proposition}\label{BMS from BEC}
		For BMSC $W$, denote $x=Z(W)$, then
		\begin{gather*} 
			\sqrt{Z_{0^n}\circ Z_{1^{m+1}} (x)}\le Z(W^{1^m0^n})\le Z_{0^n}\circ Z_{1^m}(x)\\
					Z_{1^{n-1}}\circ Z_{0^m}\circ Z_1(x)\le Z(W^{0^m1^n})\le Z_{1^n}\circ Z_{0^m}(x), \forall n\ge 1.
		\end{gather*}
	\end{proposition}
	\pf See Appendix \ref{pf of main prop}. 
	\e
	
	We obtain the first result from Proposition \ref{BMS from BEC}.
	
	\begin{theorem}\label{0r1s<1p0q}
		For any $p,n\ge 1, q,m\ge0$,
		if $0^{m}1^{n-1}\preceq_{BEC} 1^{p-1}0^{q}$, then
		$0^m1^n\preceq_Z 1^p0^q$.
	\end{theorem}
	\pf See Appendix \ref{$0^m1^n$ and $1^p0^q$}.
	\e
	
	Clearly, the well-known PO $01\preceq_Z 10$ is just an example of Theorem \ref{0r1s<1p0q}.
	Use the relationship between the  bit-error probability $P_e$ and the Bhattacharyya parameter  $Z$ in  \eqref{Z and Pe}, we have
	\begin{equation*}
	1-\sqrt{1-Z_0^n\circ Z_1^{m+1}(x)}\le 2P_e(W^{1^m0^n})\le Z_0^n\circ Z_1^m(x),
	\end{equation*}
	\begin{equation*}
	1-\sqrt{1-Z_1^{n}\circ Z_0^m\circ Z_1(x)}\le 2P_e(W^{0^m1^n})\le Z_1^n\circ Z_0^m(x),
	\end{equation*}
	where   $x=Z(W)$.
	\begin{theorem}\label{0p1s in Pe}
		For any $p\ge 1, n,q,m\ge0$, if $0^{m}1^{n}\preceq_{BEC} 1^{p-1}0^{q+1}$, then $0^m1^n\preceq_P 1^p0^q$.
	\end{theorem}
	\pf See Appendix \ref{$0^m1^n$ and $1^p0^q$}.
	\e
	
	
	Now we turn to the POs between paths with the general form. Following the proof of Proposition \ref{BMS from BEC}, we first deduce the following proposition.
	\begin{proposition}\label{general bounds for Z}
		For BMSC $W$, given path string $\alpha$ and $x=Z(W)$, then
		\begin{equation}\label{22}
		Z^{-1}_1\circ Z_{\alpha}\circ Z_1(x)\le Z(W^\alpha)\le Z_{\alpha}(x).
		\end{equation}
	\end{proposition}
	\pf  See Appendix \ref{pf of general bounds of Z}.
	\e
	
	By \eqref{22}, we  are ready to deduce  PO $\preceq_Z$ from   PO $\preceq_{BEC}$.
	\begin{theorem}\label{PO Z from cancel 1}
		For BMSC $W$, given path strings $\alpha$ and $\gamma$,
		if $\gamma\preceq_{BEC} \alpha$, then
		\begin{equation}
		\gamma 1\preceq_Z 1\alpha .
		\end{equation}
	\end{theorem}
	\pf 
	$ \gamma\preceq_{BEC}  \alpha \Rightarrow Z_{1\alpha}(x)\le Z_{1\gamma}(x), x\in[0,1].$
	Thus for any BMSC $W$ with $x=Z(W)$, according to Proposition \ref{general bounds for Z} we have
	$Z(W^{1\alpha })\le Z_{1\alpha }(x)\le Z_{1\gamma}(x)=Z_1^{-1}\circ Z_1\circ Z_\gamma\circ Z_1(x)\le Z(W^{\gamma 1}).$
	\e
	
	We provide two equivalent 
	versions of Theorem \ref{PO Z from cancel 1},   the proofs can be found  in Appendix \ref{pf Z}.
	\begin{proposition}\label{d 1}
		If $1\gamma  \preceq_{BEC}  \alpha1$, then
		$
		\gamma\preceq_Z \alpha.
		$
	\end{proposition}
	
	\begin{proposition}\label{r<_z1a}
		If $\gamma\preceq_{BEC} \alpha1$, then
		$
		\gamma\preceq_Z 1\alpha.
		$
	\end{proposition}

	Now we turn to $\preceq_P$. For convenience,  define $T(W):=2P_e(W)$. Clearly, $\alpha \preceq_P \gamma \Leftrightarrow T(W^\alpha)\ge T(W^\gamma)$ for any BMSC $W$. 
	First, we deduce the following result based on  \eqref{Z and Pe} and Proposition \ref{general bounds for Z}.
	\begin{theorem}\label{P1}
		For any BMSC $W$ and path $\alpha$, 
		\begin{equation}\label{one bound for P}
		Z_0^{-1}\circ Z_\alpha\circ Z_1(x)\le T(W^\alpha)\le Z_\alpha(x),
		\end{equation}
		with $x=Z(W)$. Furthermore, if $\gamma\preceq_{BEC}\alpha$, 
		$
		\gamma 0\preceq_P 1\alpha.
		$
	\end{theorem}
	\pf See Appendix \ref{<P 1}.
	\e
	
	Other properties of $T$ are stated as follows, the proof can be found in Appendix \ref{pf of T}.
	\begin{proposition}\label{bounds for T for one step}
		For BMSC $W$,
		\begin{equation}
		T(W^0)=Z_0(T(W));
		\end{equation}
		\begin{equation}
		Z_1(T(W))\le T(W^1)\le Z_0(T(W)).
		\end{equation}
	\end{proposition}

	\begin{proposition}\label{general bounds for Pe}
		For BMSC $W$, given path string $\alpha$ and $x=T(W)$, then
		$
		Z_{\alpha}(x)\le T(W^\alpha)\le Z_{\alpha}\circ Z_1^{-1}\circ Z_0(x).
		$
	\end{proposition}

	\begin{proposition}\label{bounds for T}
		For BMSC $W$, given path string $\alpha$ and $x=T(W)$, then
		\begin{equation}\label{01le Tle 1a}
		Z_{1\alpha}(x)\le T(W^{1\alpha})\le Z_{0\alpha} (x),
		\end{equation}
		particularly, for $\alpha=0^p1\gamma$, we have
		\begin{equation}\label{30}
		Z_{\alpha}(x)=Z_{0^{p}1\gamma}(x)\le T(W^\alpha)\le Z_{0^{p+1}\gamma}(x).
		\end{equation}
	\end{proposition}

	We can deduce PO $\preceq_P$ from $\preceq_{BEC}$ directly by \eqref{30}.
	\begin{theorem}\label{theorem of Pe}
		If $\alpha\preceq_{BEC} 0^q0\tau, q\ge 0$, then
		\begin{equation}
		\alpha\preceq_P 0^q1\tau.
		\end{equation}
	\end{theorem}

	Additionally, we show some specific rules to deduce the POs for longer paths from the shorter ones.
	\begin{theorem}\label{rule}
		For any path strings $\alpha, \gamma, \tau_1, \tau_2$,
		\begin{equation}\label{r1}
		\tau_1\preceq \tau_2, \alpha \preceq_Z  \gamma \Rightarrow \tau_1 \alpha 1^p\preceq_Z  \tau_2\gamma 1^p ,\;\;\forall p\in \mathbb{N}.
		\end{equation}
		\begin{equation}\label{r2}
		\tau_1\preceq \tau_2, \alpha \preceq_P  \gamma \Rightarrow \tau_1 \alpha 0^p\preceq_P  \tau_2\gamma 0^p ,\;\;\forall p\in \mathbb{N}.
		\end{equation}

		For any strings 	$\alpha, \tau,\gamma$ and  $p, q, r, s, t\in\mathbb{N}$, 
		\begin{equation}\label{rule2}
		0^p\alpha 1^r\preceq_Z 1^q\gamma 0^s\Rightarrow 0^p\tau \alpha 1^t 1^r\preceq_Z 1^q \tau \gamma 1^t0^s,
		\end{equation}
		\begin{equation}\label{rule3}
		0^p\alpha 1^r\preceq_P 1^q\gamma 0^s\Rightarrow 0^p\tau \alpha 0^t 1^r\preceq_P 1^q \tau \gamma 0^t0^s.
		\end{equation}
	\end{theorem}
	\pf  See Appendix \ref{pf of rules}.
	\e

	In the end of this part, we show an application of Proposition \ref{d 1} and Theorem \ref{theorem of Pe}.
	The reliability of a path $\alpha$   depends on the number and the positions of 1s. In the following, we provide a sufficient condition on the numbers of 1s and 0s, under which two paths $\alpha$ and $\gamma$ are comparable regardless of the order of 0 and 1.    Here we use $n_i(\alpha)$ to denote the number of $i$s in $\alpha,i=0,1$.

	\begin{corollary}\label{Co3.5}
		Given a path string $\alpha$ with   $|\alpha|=n$, then:
		\begin{enumerate}
			\item If
			$
			n_0(\alpha)\le \log_2(n-\log_2 (n)),
			$
			we conclude that for any length-$n$ path $\gamma$, let $\tau=\gamma_1\cdots\gamma_{n-1}$, if $n_0(\tau)\ge n_1(\alpha)$ and $n_1(\tau)\le n_0(\alpha)-1,$ then
			\begin{equation}
			\gamma\preceq_{Z} \alpha.
			\end{equation}

			\item If
			$
			n_0(\alpha)\le \log_2(n-\log_2 (n))-1,
			$
			we conclude that for any $n$-length $\gamma$, if  $n_0(\gamma)\ge n_1(\alpha)-1$ and  $n_1(\gamma)\le n_0(\alpha)+1$, then 
			\begin{equation}
			\gamma\preceq_{P} \alpha.
			\end{equation}
		\end{enumerate} 
	\end{corollary}
	\pf See Appendix \ref{pf of CO3.5}.
	\e
	%
	

	\subsection{Examples}
	In this part, we provide  specific POs derived from Theorem \ref{PO Z from cancel 1}-\ref{rule}, which cannot be deduced from the existing POs.
	\begin{enumerate}
		\item
		Examples of Theorem \ref{PO Z from cancel 1}
		\begin{enumerate}
			
			\item $11001\preceq_Z 10111$, since $1100\preceq_{BEC}0111$
			
			\item $110001\preceq_Z 101101$, since $11000\preceq_{BEC} 01101$.
			
			\item If $(1-1/2^{2^m})^{2^n}\le 1/2$, then $1^{m}0^n\preceq_Z 10^m1^{n-1}.$ Particularly, for any positive integer $k$,
			$
			1^{k}0^{2^k}\preceq_Z 10^k 1^{2^k-1}.
			$
		\end{enumerate}
		
		\item
		Examples of Proposition \ref{d 1}:
		\begin{enumerate}
			\item $100\preceq_Z 011$, since $1100\preceq_{BEC}0111$.

			\item $1010\preceq_Z 0111$, since $11010\preceq_{BEC} 01111$.

			\item If $(1-1/2^{2^m})^{2^n}\le 1/2$, then $1^{m-1}0^n\preceq_Z 0^m1^{n-1}.$ Particularly, for any positive integer $k$,
			$
			1^{k-1}0^{2^k}\preceq_Z 0^k 1^{2^k-1}.
			$
		\end{enumerate}

		\item Examples for Theorem \ref{P1}:
		\begin{enumerate}
			\item 	  $11000\preceq_P 10111$, since $1100\preceq_{BEC}0111$.
			
			\item 	  $110100\preceq_P 101111$, since $11010\preceq_{BEC} 01111$.
			
			\item 	    If $(1-1/2^{2^m})^{2^n}\le 1/2$, then $1^{m}0^{n+1}\preceq_P 10^m1^{n}.$ Particularly, for any positive integer $k$,
			$
			1^{k}0^{2^k+1}\preceq_P 10^k 1^{2^k}.
			$
			
		\end{enumerate}

		\item Examples of Theorem \ref{theorem of Pe}:
		\begin{enumerate}
			
			\item $1000\preceq_P 0111$, since $1000\preceq_{BEC} 0011$.
			
			\item $10010\preceq_P 01111$, since $10010\preceq_{BEC} 00111$.

			
			\item If $(1-1/2^{2^m})^{2^n}\le 1/2$, then $1^{m}0^n\preceq_P 0^{m-1}1^{n+1}.$ Particularly, for any positive integer $k$,
			$
			1^{k}0^{2^k}\preceq_P 0^{k-1} 1^{2^k+1}.
			$
		\end{enumerate}
		
		%
		%
		%

		\item Examples for Theorem \ref{rule}:
		\begin{enumerate}
			\item $1010(100)11\preceq_Z 1010(011)11$, since $100\preceq_Z 011$.
			
			\item $01(100001)\preceq_P 10(011110)$, since $100001\preceq_{BEC}001110\Rightarrow100001\preceq_P 011110$.
			
			\item $0(01)11000(11)1\preceq_Z 1(01)00111(11)0$, since $011000\preceq_{BEC} 001110\Rightarrow 0110001\preceq_Z 1001110.$

			\item  $0(0101)110000(00)1\preceq_P 1(0101)001111(00)0$,  since $01100001\preceq_P 10011110$.
		\end{enumerate}

		\item Bounds of $\beta$ in $\beta$-expansion:
		\begin{enumerate}
			\item 	Since $01\preceq10$, $\beta\ge 1$.
			

			\item Since $1100\preceq_Z 1011$, $0<\beta\le \frac{1+\sqrt{5}}{2}$.
			
			\item Since $1010\preceq_Z 0111$, $0<\beta\le -\frac{1}{3}+\sqrt[3]{\frac{29}{54}+\sqrt{\left( \frac{29}{54}\right) ^2-\frac{1}{729}}}+\sqrt[3]{\frac{29}{54}-\sqrt{\left( \frac{29}{54}\right) ^2-\frac{1}{729}}}.$
			
			
			\item We find that $\beta\in[1,2^{0.30}]$ meet all the POs $\preceq_Z$ when $n\le 10$.
			
		\end{enumerate}
		
	\end{enumerate}
	
	\section{Simulation}\label{section4}
	When $n=10$, there are $C_{1024}^2=523776$ path pairs in total. Based on the known channel degradation results stated in Proposition \ref{known PO in BMS}, there are 328155 pairs of POs $\preceq$, denoted as $P_k$. According to Theorem \ref{PO Z from cancel 1}, we discover 378796 pairs of PO $\preceq_Z$, denoted as $P_b$. We observe that $|P_b\backslash P_k|=50641$, indicating that there are 50641 pairs of POs that cannot be deduced from the known POs. The proportions of the existing POs, the newly discovered POs and the unknown pairs   are shown in Fig. 1.

	Fig. 2 illustrates the SC decoding performance of polar codes with various values of $\beta$. We compare the PO set $P_b$ with $\preceq_\beta$. It is observed that when $\log_2(\beta)$ lies within the range of $[0, 0.30]$, there are no conflicts between $P_b$ and the $\beta$-expansion approach. Additionally, we use $P_b$ to modify the path ordering when $\log_2\beta>0.30$, the modified polar codes demonstrate improved error-correcting performance.

	Violating new POs may result in a performance penalty. As an example, let
	$
	P_u:=\{(719, 250), (840,372), (907,466),(909,690),(921,482) \}\subset P_b\backslash P_k,
	$
	where the former bits are superior to the latter ones in all pairs. We compare two  information sets: $\mathcal{A}_{5G}$ and $\mathcal{A}_1$ with code length $N=1024$, where $\mathcal{A}_{5G}$ is constructed by reliability sequence in 5G \cite{b21} and
	$
	\mathcal{A}_1:=(\mathcal{A}_{5G}\backslash \{719,840,907,909,921\})\bigcup \{250,372,466,482,690\}.
	$
	Fig. 3 presents the performance comparison between the two polar codes mentioned above. It is observed that there is a penalty of 0.25 dB for $\mathcal{A}_1$ compared to $\mathcal{A}_{5G}$.
	
		\begin{figure}[!t]
		\centering
		\includegraphics[width=5.5cm,height=4cm]{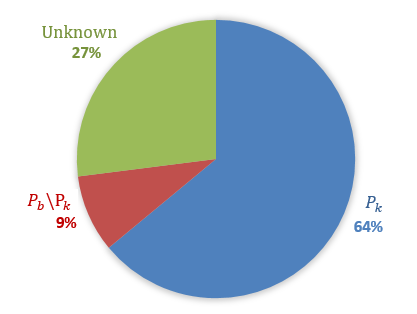}
		\caption{The proportions of the existing POs, the newly discovered POs and the unknown pairs   when $n=10$.}
	\end{figure}
	\begin{figure}[!t]
		\centering
		
		\includegraphics[width=7cm,height=5cm]{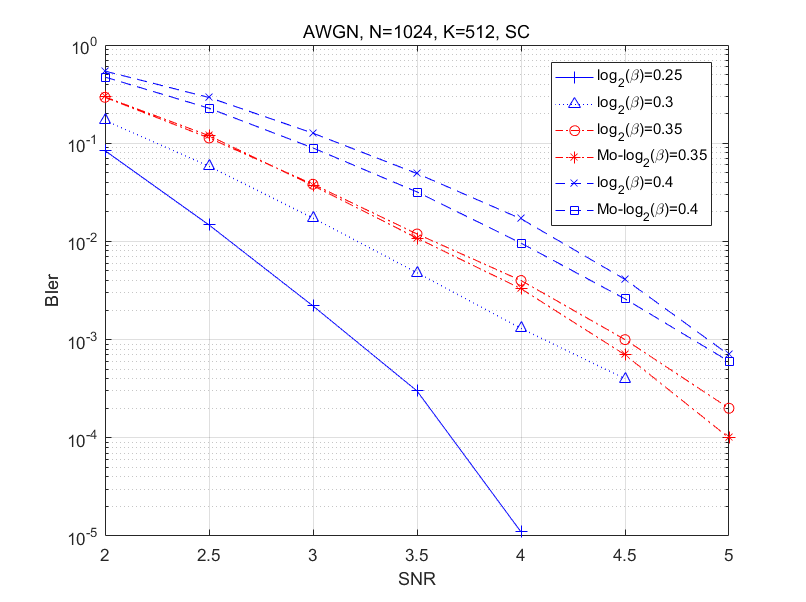}
		\caption{SC decoding performance under different $\beta$, where    curves with the prefix 'MO' are the modification versions according to $P_b$.}
	\end{figure}
	\begin{figure}[!t]
		\centering
		\includegraphics[width=7.5cm,height=5.3cm]{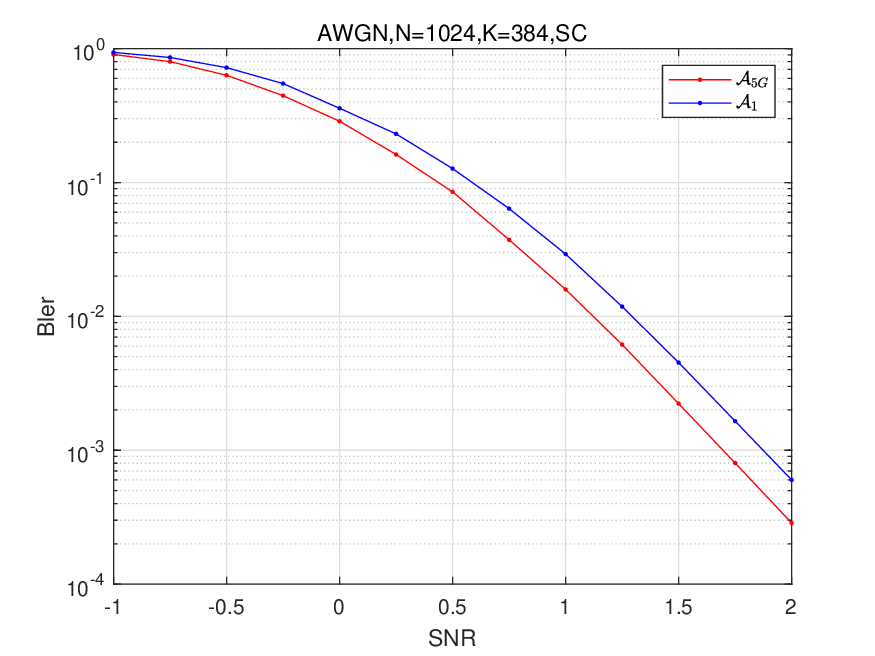}
		\caption{SC decoding performance with different information sets $\mathcal{A}_{5G}$ and $\mathcal{A}_1$. }
	\end{figure}

	\section{Conclusion}\label{section5}
	In this paper, we introduce two POs for BMSC based on the Bhattacharyya parameter and the bit-error probability, which are  more computable compared with channel degradation. As summarized in Table \ref{table 1}, we discover new POs based on the results in BEC. We also derive upper bounds for $\beta$ by certain examples derived from our results.
	
	However, from the practical point of view, we only need to compare the synthetic channels within the target SNR regime. Therefore, the stringent nature of the existing POs narrows their scope of application. We leave the studies on   practical POs for future research.

	\begin{appendix}
		\subsection{Proof of Proposition \ref{BMS from BEC}}\label{pf of main prop}
		According to   Proposition \ref{bound functions}, Lemma \ref{functions} and the fact that functions $Z_0(x), Z_1(x), L_0(x)$ are all increasing in $[0,1]$, we can immediately write that 
		\begin{equation}
		L_0^{(n)}\circ Z_{1^m}(x)\le Z(W^{1^m0^n})\le Z_{0^n}\circ Z_{1^m}(x);
		\end{equation}
		\begin{equation}
		Z_{1^n}\circ L_0^{(m)}(x)\le Z(W^{0^m1^n})\le Z_{1^n}\circ Z_{0^m}(x).
		\end{equation}
		Recall that Lemma \ref{functions}  implies that
		\begin{equation}
		L_0^{(k)}=Z_1^{-1}\circ Z_{0^k}\circ Z_1(x),\;\;\forall x\in[0,1], k=0,1,2,\cdots,
		\end{equation}
		the proof is completed.
		\e

		\subsection{Proof of the Results for $0^m1^n$ and $1^p0^q$}\label{$0^m1^n$ and $1^p0^q$}
		\textbf{Proof of Theorem \ref{0r1s<1p0q}.} 
		$0^{m}1^{n-1}\preceq_{BEC} 1^{p-1}0^{q}\Rightarrow  Z_{0^m1^{n-1}}(x)\ge Z_{1^{p-1}0^q}(x), \forall x\in [0,1]\Rightarrow  Z_{0^m1^{n-1}}(x^2)\ge Z_{1^{p-1}0^q}(x^2), \forall x\in [0,1]\Rightarrow  Z_{10^m1^{n-1}}(x)\ge Z_{1^{p}0^q}(x), \forall x\in [0,1].$
		Thus for any BMSC $W$  with $x=Z(W)$, by Proposition \ref{BMS from BEC}  we have
		\begin{equation}
		Z(W^{ 1^p0^q})\le  Z_{1^{p}0^q}(x)\le Z_{10^m1^{n-1}}(x)\le Z(W^{0^m1^n}).
		\end{equation}
		\e

		\textbf{Proof of Theorem \ref{0p1s in Pe}.}
		Note that $Z_0(1-\sqrt{1-Z_1^{n}\circ Z_0^m\circ Z_1(x)})=Z_1^{n}\circ Z_0^m\circ Z_1(x)=Z_1^{n}\circ Z_0^m(x^2)$, and $Z_0(Z_0^q\circ Z_1^p(x))=Z_0^{q+1}\circ Z_1^{p-1}(x^2)$. We complete the proof since $Z_0(x)$ is non-decreasing on $[0,1]$.
		\e

			\subsection{Proof of Proposition \ref{general bounds for Z}}\label{pf of general bounds of Z}
		Set $\alpha=1^{q_1}0^{p_1}1^{q_2}0^{p_2}\cdots 1^{q_m}0^{p_m}$,  thanks to Proposition \ref{BMS from BEC}, we have
		\begin{equation}
		\begin{aligned}
		&Z^{-1}_1\circ Z_{0^{p_m}}\circ Z_{1^{q_m}}\circ Z_{0^{p_{m-1}}}\circ\cdots\circ Z_{1^{q_1}}\circ Z_1(x)\\
		&\le Z(W^\alpha)\\
		&\le Z_{0^{p_m}}\circ Z_{1^{q_m}}\circ Z_{0^{p_{m-1}}}\circ Z_{1^{q_{m-1}}}\circ\cdots\circ Z_{0^{p_1}}\circ Z_{1^{q_1}}(x).
		\end{aligned}
		\end{equation}
		\e

		\subsection{Proofs of the Equivalent Versions of Theorem \ref{PO Z from cancel 1}}\label{pf Z}
		\textbf{Proof of Proposition \ref{d 1}.} $1\gamma  \preceq_{BEC}  \alpha 1 \Rightarrow Z_{\alpha 1}(x)\le Z_{1\gamma}(x), x\in[0,1]\Rightarrow Z_{\alpha }(x)\le Z_{1}^{-1}\circ Z_{1\gamma}(x), x\in[0,1].$
		Thus for any BMSC $W$ with $x=Z(W)$, according to Proposition \ref{general bounds for Z} we have
		\begin{equation}
		Z(W^\alpha)\le Z_{\alpha }(x)\le Z_{1}^{-1}\circ Z_{ \gamma}\circ Z_1(x)\le Z(W^\gamma).
		\end{equation} 
		\e
		
		\textbf{Proof of Proposition \ref{r<_z1a}.} $\gamma\preceq_{BEC} \alpha1$  means $1\gamma\preceq_{BEC} 1\alpha 1$, then by Proposition \ref{d 1}, $\gamma\preceq_Z 1\alpha$.
		\e
		
		We also provide the proof for the equivalence between Theorem \ref{PO Z from cancel 1}, Proposition \ref{d 1} and Proposition  \ref{r<_z1a}.
		
		\begin{enumerate}
			\item Theorem \ref{PO Z from cancel 1} $\Rightarrow$ Proposition \ref{d 1}. By the proof of Theorem \ref{PO Z from cancel 1}, we conclude that for any BMSC $W$ with $x=Z(W)$,
			$
			1\gamma\preceq_{BEC}\alpha 1 
			$ means the lower bound of $Z(W^{1\gamma 1})$ in \eqref{22} is larger than the upper bound of $Z(W^{1\alpha 1})$ in \eqref{22}, \textit{i.e.}
			\begin{equation*}
			Z_{1\alpha 1}(x)\le Z_{11\gamma}(x),
			\end{equation*}
			and thus by \eqref{22}, we have
			\begin{equation*}
			Z(W^\alpha)\le Z_{\alpha }(x)\le Z_1^{-1}\circ Z_{1\gamma}(x)\le Z(W^\gamma),
			\end{equation*}
			\textit{i.e.} $\gamma\preceq_Z \alpha$.
			
			\item Proposition \ref{d 1} $\Rightarrow$ Proposition \ref{r<_z1a}. See the proof of Proposition \ref{r<_z1a}.
			
			\item Proposition \ref{r<_z1a} $\Rightarrow$ Theorem \ref{PO Z from cancel 1}. $\gamma\preceq_{BEC}\alpha \Rightarrow \gamma 1\preceq_{BEC} \alpha 1\Rightarrow \gamma 1\preceq_{Z} 1\alpha.$  
		\end{enumerate}
		\e
		
		\subsection{Proof of Theorem \ref{P1}}\label{<P 1}
		\eqref{one bound for P} is from \eqref{Z and Pe} and Proposition \ref{general bounds for Z},   thus for any BMSC $W$ with $x=Z(W)$,
		\begin{align*}
		&\gamma\preceq_{BEC}\alpha\Rightarrow 1\gamma 0\preceq_{BEC}1\alpha 0\\
		&\Rightarrow Z_{1\alpha }(x)\le Z_{1\gamma }(x),\;\;\forall x\in[0,1]\\
		&\Rightarrow Z_{1\alpha }(x)\le Z_0^{-1}\circ Z_{\gamma 0}\circ Z_1(x),\;\;\forall x\in[0,1],
		\end{align*}
		then  the proof is completed by \eqref{one bound for P}.
		\e

		\subsection{Proofs of the Properties for $T(W)$}\label{pf of T}
		\textbf{Proof of Proposition \ref{bounds for T for one step}.} 
		We only prove the second bounds. In fact, by \eqref{Z and Pe},
		\begin{align}
		&Z_1(T(W))=4P_e(W)^2\le 2P_e(W^1)\le Z(W^1)\\
		&=Z(W)^2\le 1-(1-2P_e(W))^2=Z_0(T(W)).
		\end{align}
		\e
		
		\textbf{Proof of Proposition \ref{general bounds for Pe}.}  
		$Z_{\alpha}(x)\le T(W^\alpha)$ comes from Proposition \ref{bounds for T for one step}. In addition, by \eqref{Z and Pe},
		\begin{align}
		T(W^\alpha)\le Z(W^\alpha)\le Z_{\alpha}(Z(W))
		\le Z_{\alpha}(\sqrt{Z_0(x)}).
		\end{align}
		\e

		\textbf{Proof of Proposition \ref{bounds for T}.}   
		\eqref{01le Tle 1a} can be deduced from Proposition \ref{general bounds for Pe} directly, and
		\begin{align}
		&Z_{\alpha}(x)=Z_{0^{p}1\gamma}(x)=Z_{1\gamma}(T(W^{0^p}))\le T((W^{0^p})^{1\gamma})\\
		&=T(W^\alpha)\le Z_{0\gamma}(T(W^{0^p}))=Z_{0^{p+1}\gamma}(x).
		\end{align}
		\e
		
		\subsection{Proof of Theorem \ref{rule}}\label{pf of rules}
		Since $\tau_1\preceq \tau_2$, we have $\tau_1\gamma\preceq \tau_2\gamma$. Then for any BMSC $W$, let $V_1:=W^{\tau_1}$ and $V_2:=W^{\tau_2}$, then
		\begin{equation*}
		\begin{aligned}
		&Z(W^{\tau_1\alpha 1^p})=Z_{1^p}(Z(V_1^{ \alpha }))\ge Z_{1^p}(Z(V_1^{ \gamma }))\\
		&=Z_{1^p}(Z(W^{ \tau_1\gamma }))
		\ge Z_{1^p}(Z(W^{ \tau_2\gamma }))
		=Z(W^{\tau_2\gamma 1^p}),
		\end{aligned}
		\end{equation*}
		\begin{equation*}
		\begin{aligned}
		&T(W^{\tau_1\alpha 0^p})=Z_{0^p}(T(V_1^{ \alpha }))
		\ge Z_{0^p}(T(V_1^{ \gamma }))\\
		&=Z_{0^p}(T(W^{ \tau_1\gamma }))
		\ge Z_{0^p}(T(W^{ \tau_2\gamma }))
		=T(W^{\tau_2\gamma 0^p}).
		\end{aligned}
		\end{equation*}
		The proofs of \eqref{r1} and \eqref{r2} are completed. 
		
		Now we turn to \eqref{rule2}. With   $t\in\mathbb{N}$ fixed, we consider  $\tau=0$ or 1. When $\tau=0$, by \eqref{r1},
		\begin{align*}
		&0^p\alpha 1^r\preceq_Z 1^q\gamma 0^s \Rightarrow 00^p\alpha 1^r1^t\preceq_Z 01^q\gamma 0^s1^t\\
		&\Rightarrow  0^p0\alpha 1^t1^r\preceq_Z 01^q\gamma 0^s1^t.
		\end{align*}
		Thanks to Proposition \ref{known PO in BMS}, we have 
		$$01^q\gamma 0^s1^t\preceq  1^q0\gamma1^t 0^s,$$
		thus we conclude $0^p0 \alpha 1^t 1^r\preceq_Z 1^q 0 \gamma 1^t0^s$.
		When $\tau=1$, by \eqref{r1} and Proposition \ref{known PO in BMS}, we have
		\begin{align*}
		&0^p\alpha 1^r\preceq_Z 1^q\gamma 0^s \Rightarrow 10^p\alpha 1^r1^t\preceq_Z 11^q\gamma 0^s1^t\\
		&\Rightarrow  0^p1\alpha 1^t1^r\preceq_Z  1^q1\gamma 1^t0^s.
		\end{align*}
		Now we prove \eqref{rule2}   by induction on the length of $|\tau|$. Indeed, assuming that \eqref{rule2} is true for $|\tau|=k$, then for any string $\tau$ with $|\tau|=k+1$,   $\alpha'=\tau_2\tau_3\cdots\tau_{k+1}\alpha1^t$, and $\gamma'=\tau_2\tau_3\cdots\tau_{k+1}\gamma1^t$,
		\begin{align*}
		&0^p\alpha 1^r\preceq_Z 1^q\gamma 0^s\Rightarrow 0^p\tau_2\tau_3\cdots\tau_{k+1}\alpha 1^t 1^r\preceq_Z 1^q\tau_2\tau_3\cdots\tau_{k+1}\gamma 1^t0^s\\
		&\Rightarrow 0^p\alpha' 1^r\preceq_Z 1^q\gamma' 0^s\Rightarrow 0^p\tau_1\alpha' 1^r\preceq_Z 1^q\tau_1\gamma' 0^s\\
		&\Rightarrow 0^p\tau \alpha 1^{t} 1^r\preceq_Z 1^q \tau \gamma 1^{t}0^s.
		\end{align*}
		We can also prove \eqref{rule3} in a similar manner.
		\e

		\subsection{Proof of Corollary \ref{Co3.5} }\label{pf of CO3.5}
		We first provide a lemma related to PO $\preceq_{BEC}$.
		\begin{lemma}\label{BECBSC}
			Given a path string $\alpha$,  then for any $\gamma$ with $|\gamma|=n, n_0(\gamma)\ge n_1(\alpha), n_1(\gamma)\le n_0(\alpha)$,
			
			$n_0(\alpha)\le log_2 n_1(\alpha)    \Rightarrow \gamma\preceq_{BEC}\alpha.$
			
		\end{lemma}
		\pf  Denote $k=n_0(\alpha)$,

		\begin{align*}
		&2^{k}+k\le n \Rightarrow -\frac{2^{n-k}}{2^{2^k}}\le -1\Rightarrow 2^{n-k}\log_2\left( 1-\frac{1}{2^{2^k}}\right) \le -1\\
		&\Rightarrow \left( 1-\frac{1}{2^{2^k}}\right)^{2^{n-k}}\le \frac{1}{2}\Rightarrow 	(1-1/2^{2^{n_0(\alpha)}})^{2^{n-n_0(\alpha)}}\le 1/2.\\
		\end{align*}
		The proof completes from   \cite[Theorem 1]{lin2023optimal},   and the fact that
		$
		0^{n_0(\alpha)}1^{n-n_0(\alpha)}\preceq  \alpha\preceq 1^{n-n_0(\alpha)}	0^{n_0(\alpha)}.
		$
		
		\e
		
		%
		%
		%

		Lemma \ref{BECBSC} implies that when $n_0(\alpha)\le \log_2(n-\log_2n),$ where $|\alpha|=n$, we have 
		\begin{equation}\label{base1}
		\bar{\alpha}\preceq_{BEC} 1^{n_0(\alpha)}0^{n_1(\alpha)}\preceq_{BEC} 0^{n_0(\alpha)}1^{n_1(\alpha)}\preceq_{BEC} \alpha.
		\end{equation}
		Combining \eqref{base1}, Proposition \ref{d 1} and Theorem \ref{theorem of Pe}, we provide the proof of Corollary \ref{Co3.5}.

		\textbf{Proof of Corollary \ref{Co3.5}.} Without loss of generality, assume $n>1$.
		\begin{enumerate}
			\item Since $n_0(\alpha)\le \log_2(n-\log_2(n))$, by \eqref{base1} we have
			\begin{equation*}
			1^{n_0(\alpha)}0^{n_1(\alpha)}\preceq_{BEC}   \alpha,
			\end{equation*}
			thus 
			\begin{equation*}
			1^{n_0(\alpha)}0^{n_1(\alpha)}1\preceq_{BEC} 0^{n_0(\alpha)}1^{n_1(\alpha)}1,
			\end{equation*}
			by Proposition \ref{d 1},
			\begin{equation*}
			1^{n_0(\alpha)-1}0^{n_1(\alpha)}1\preceq_Z 0^{n_0(\alpha)}1^{n_1(\alpha)}\preceq \alpha,
			\end{equation*}
			thus
			\begin{equation*}
			\begin{aligned}
			&\gamma=\tau\gamma_n\preceq 1^{n_1(\tau)}0^{n_0(\tau)}\gamma_n\\
			&\preceq 1^{n_0(\alpha)-1}0^{n_1(\alpha)}\gamma_n\preceq 1^{n_0(\alpha)-1}0^{n_1(\alpha)}1\preceq_Z \alpha. 
			\end{aligned}
			\end{equation*}
			
			\item The assumptions imply that
			\begin{equation*}
			\begin{aligned}
			&1^{n_0(\alpha)+1}0^{n_1(\alpha)-1}\preceq_{BEC}0^{n_0(\alpha)+1}1^{n_1(\alpha)-1},
			\end{aligned}
			\end{equation*}
			thus by Theorem \ref{theorem of Pe},
			$$1^{n_0(\alpha)+1}0^{n_1(\alpha)-1}\preceq_{P}0^{n_0(\alpha)}1^{n_1(\alpha)}\preceq\alpha.$$
			Then 
			\begin{equation*}
			\begin{aligned}
			&\gamma\preceq 1^{n_1(\gamma)}0^{n_0(\gamma)}\preceq 1^{n_0(\alpha)+1}0^{n_1(\alpha)-1}\\
			&\preceq_{P}0^{n_0(\alpha)}1^{n_1(\alpha)}\preceq \alpha.
			\end{aligned}
			\end{equation*}
		\end{enumerate}
		
		\e
		
	\end{appendix}
	

\begin{thebibliography}{10}
		
		\bibitem{arikan2008}
		E. Arikan.
		\newblock Channel polarization: A method for constructing capacity-achieving
		codes.
		\newblock In {\em 2008 IEEE International Symposium on Information Theory},
		pages 1173--1177, 2008.
		
		\bibitem{densityTanaka}
		R. Mori and T. Tanaka.
		\newblock Performance of polar codes with the construction using density
		evolution.
		\newblock In {\em IEEE Communications Letters}, 13(7):519--521, 2009.\
		
		\bibitem{b11} P. Trifonov. 
		\newblock Efficient Design and Decoding of Polar Codes.
		\newblock In {\em IEEE Transactions on Communications}, 60(11):3221-3227, 2012.
		
		\bibitem{beta-expansion}
		G. He, J.-C. Belfiore, I. Land, G.-H. Yang, X.-C. Liu,
		Y. Chen, R. Li, J. Wang, Y.-Q. Ge, R. Zhang, and W. Tong.
		\newblock Beta-expansion: A theoretical framework for fast and recursive
		construction of polar codes.
		\newblock In {\em 2017 IEEE Global Communications Conference},
		pages 1--6, 2017.
		
		\bibitem{2011How}
		I.Tal and A. Vardy.
		\newblock How to construct polar codes.
		\newblock In {\em IEEE Transactions on Information Theory}, 59(10):6562--6582, 2011.
		
		
		\bibitem{2021greedy}
		J. Muramatsu.
		\newblock Binary polar codes based on bit error probability.
		\newblock In {\em 2022 IEEE International Symposium on Information Theory},
		pages 2148--2153, 2022.
		
						\bibitem{01<10}
		M. Bardet, V. Dragoi, A. Otmani, and J. -P. Tillich
		\newblock Algebraic properties of polar codes from a new polynomial formalism.
		\newblock In {\em 2016 IEEE International Symposium on Information Theory},
		pages 230--234, 2016.
		
		\bibitem{2016PO}
		C. Schurch.
		\newblock A partial order for the synthesized channels of a polar code.
		\newblock In {\em 2016 IEEE International Symposium on Information Theory}, 
		pages 220--224, 2016.
		
		\bibitem{Wu2019}
		W. Wu and P. H. Siegel.
		\newblock Generalized partial orders for polar code bit-channels.
		\newblock In {\em IEEE Transactions on Information Theory}, 65(11):7114--7130,
		2019.
		
		
		\bibitem{lin2023optimal}
		T.-C. Lin and H.-P. Wang.
		\newblock Optimal self-dual inequalities to order polarized BEC.
		\newblock In {\em 2023 IEEE International Symposium on Information Theory},
		pages 1550--1555, 2023.
		
		
		\bibitem{wang2023density}
		H.-P. Wang and C.-W. Chin.
		\newblock Density devolution for ordering synthetic channels.
		\newblock In {\em 2023 IEEE International Symposium on Information Theory},
		pages 1544--1549, 2023.
		
		\bibitem{wang2023fast}
		H.-P. Wang and V.-F. Drăgoi.
		\newblock Fast methods for ranking synthetic BEC.
		\newblock In {\em 2023 IEEE International Symposium on Information Theory},
		pages 1562--1567, 2023.
		
		
		
		
		\bibitem{2008Modern}
		T. Richardson and R. Urbanke.
		\newblock Modern coding theory.
		\newblock {\em Cambridge University Press}, 2008.
		
		
		
		
		\bibitem{relaxed2017}
		M. El-Khamy, H. Mahdavifar, G. Feygin, J. Lee, and I.
		Kang.
		\newblock Relaxed polar codes.
		\newblock In {\em IEEE Transactions on Information Theory}, 63(4):1986--2000,
		2017.
		
		
		
		
		
		
		
		
		
		
		
		
		
		
		
		\bibitem{b21} 3GPP, "NR; Multiplexing and channel coding", 3GPP TS 38.212,15.5.0, Mar. 2019.
		
	\end{thebibliography}
\end{document}